\documentclass[12pt,a4paper]{article}
\usepackage{amsmath}
\input epsf
\usepackage[dvips]{graphicx}
\usepackage{times}

\begin{document}
\begin{titlepage}
\title{Unitarity: confinement and collective effects in hadron interactions}
\author{ S.M. Troshin,  N.E. Tyurin\\[1ex] \small\it Institute
\small\it for High Energy Physics,\\\small\it Protvino, Moscow Region, 142281 Russia}
\date{}
\maketitle
\begin{abstract}
We discuss how saturation of unitarity would change phase structure of hadronic matter at very high 
temperatures emphasizing the role of the vacuum state with spontaneously broken chiral symmetry.
\end{abstract}
\vfill
\end{titlepage}
\section*{Introduction}

It is widely known, that the main fundamental
problems of  QCD  are related to confinement and
spontaneous chiral symmetry breaking phenomena. Those phenomena
are associated with collective, coherent interactions of quarks and gluons,
and result in formation of the asymptotic states, which are the
colorless, experimentally observable particles. Owing to experimental efforts 
during recent decades it has
become evident that the coherent collective dynamics survives at  high energies.

The hypothesis 
on the completeness of the set of asymptotic states plays an important role  (cf. \cite{bart,rlst})
and leads, e.g. due to  unitarity of scattering matrix to the optical theorem relating the total cross-section
with  the forward elastic scattering amplitude.
Thus, the general principles  play a guiding  role in  hadron interaction studies,
 and, in particular, unitarity which regulates the
relative strength of elastic and inelastic processes is  the most
significant one. It is important to note here again that unitarity is formulated
for the asymptotic colorless hadron on-mass shell states and is not directly connected
to the fundamental fields of QCD --- color filelds of quarks and gluons.

The Hilbert space corresponds to colorless hadron states, it is constructed using vectors obtained by 
acting  with the relevant creation operators on the physical vacuum. The state of  physical vacuum 
as it is defined in the axiomatic field theory
is a state without particles, annihilation operator when acts on it produces zero. It is the state of lowest energy
and  invariant under Lorentz transformations. 
Nowadays it is  accepted that this vacuum state  is not
unique. 
Indeed, colored  current quarks and gluons
are the degrees of freedom related to the perturbative vacuum which is different from physical one. 
According to the confinement property
of QCD, isolated colored objects cannot exist in the physical vacuum. Transition from physical vacuum 
to the perturbative one occurs in the  process of deconfinement and results in quark-gluon plasma
formation, i.e.   gaseous state of free colored quarks and gluons.   It is clear that
hadrons  and free quarks and gluons  cannot coexist  together since
they  live in different vacua \cite{mingmei} and  there is no room for objects like
quark-proton scattering amplitude. 
This fact provides important restrictions on the possible
mechanisms of deconfinement.

The picture described above  is commonly accepted in the theory of strong interactions. The main
ingredient here is the assumption on the same scale of transitions confiment-deconfinement and chiral restoration. 
This assumption has a theoretical ground in some of lattice calculations (cf. e.g. \cite{born}).  

However, it is often assumed
that the scales relevant to confinement and chiral symmetry breaking are different \cite{manoh}, scale of confinement 
is $\Lambda_{QCD} = 100-300$ MeV while chiral symmetry breaking scale --- $\Lambda_\chi\simeq 1$ GeV. 
Thus, in the range 
between these two scales the matter is in a deconfined state but chiral symmetry is spontaneously broken there. 
In the line with this picture, which can be treated as a posteriori justification, long time ago, in the pre-QCD era, 
it was supposed that hadrons have a simple structure and
nonrelativistic quark model has been commonly adopted. During recent time  such a model has evolved and obtained much more
solid theoretical ground \cite{manoh,kaplan,diak}. As it will be discussed further, one can assume existence inside the
hadron  of the third (nonperturbative) vacuum state with colored constituent quarks
and pions as relevant degrees of freedom.

In this note we would like to consider how unitarity can  constrain dynamics 
of the confined objects. We look into a possible mechanism of deconfinement assuming existence of a nonperturbative 
vacuum in addition to  perturbative and physical ones and study the role of unitarity saturation in the 
process of deconfinement. It will be shown that saturation of unitarity is acting as a confinement restoration process.
This is not surprising, since unitarity implies completeness of asymptotic colourless states.
In this case it leads to appearance of a new state of a hadron matter at superhigh temperatures.

\section{Nonperturbative vacuum and effective degrees of freedom}

The origin of the nonperturbative vacuum and relevant effective degrees of fredom are related to
the mechanism of spontaneous chiral symmetry breaking ($\chi$SB) in QCD \cite{bjorken},
 which  leads
to  generation of quark masses and appearance of quark condensates. This mechanism describes
transition of the current into  constituent quarks.
Massive  constituent quarks appear  as quasiparticles, i.e. current quarks and
the surrounding  clouds  of quark--antiquark pairs.  

Collective excitations of the condensate are the Goldstone bosons,
and the constituent quarks interact with each other via exchange
of the Goldstone bosons; this interaction is mainly due to pion field.
Pions themselves are the bound states of massive
quarks.
Thus, the effective unteraction of constituent quarks proceeds through  exchange
of Goldstone bosons. Constituent quark interactions with Goldsone bosons  is strong and could have 
  the following form  \cite{diak}:
 \begin{equation}
{\cal{L}}_I=\bar Q[i\partial\hspace{-2.5mm}/-M\exp(i\gamma_5\pi^A\lambda^A/F_\pi)]Q,\quad \pi^A=\pi,K,\eta.
\end{equation}
For simplicity, in what follows we will refer to  pions only, denoting by this 
generic word all Goldstone bosons, i.e. pions themselves, kaons
 and $\eta$-mesons.

Thus, we will assume that  vacuum state $V_{pt}$ has a perturbative nature at short distances
with current quarks and gluons as degrees of freedom, at large distances the physical vacuum state 
$V_{ph}$ has relevant colorless hadrons as degrees of freedom, and inside a hadron  the vacuum $V_{np}$
has a nonperturbative origin with constituent quarks and Goldstone pions being relevant degrees of freedom.
We suppose the picture of a hadron consisting of constituent quarks embedded
 into quark condensate and interacting with pions which have a  dual role: Goldstone and physical particles. 

There are different approaches to the deconfinement dynamics, e.g. the deconfinement mechanism  can be
formulated in terms of percolation theory. It was recently \cite{satz,satz2}  proposed to use
it as a candidate for the mechanism of deconfinement in the form of  analytical
crossover (without first and second order phase transitions).  This form of deconfinement 
was found in the experimental studies at RHIC.
 Evidently such  purely geometrical approach should be amended by  a 
dynamical mechanism and indeed
color dynamics of deconfinement due to formation
of molecular-like aggregations was proposed in \cite{mingmei}. The vacuum 
inside the hadron was taken to be a perturbative one and quark interactions have
origin in the color dynamics. It seems, however, that for crossover nature of deconfinement dynamics 
 it is more natural to
expect transition $V_{ph}\to V_{np}$ instead of transition $V_{ph}\to V_{pt}$. 
Indeed, using effective quark-pion interaction inside hadron and hadron-pion interaction outside
hadron, we have a pion field as an universal interaction agent for both confined and deconfined states
 and this could serve as a natural explanation of deconfinement as a cross-over transition.

Experimentally deconfined state of matter has been discovered  
at RHIC where the highest values of energy and
density have been reached. This deconfined
state appears to be strongly interacting collective state
with properties of  perfect liquid.  It is interesting to note that  phase transition from parton
gas to liquid could explain  saturation phenomena in deep inelastic processes \cite{jenk}.  
The importance of the
experimental discoveries at RHIC is that the matter is strongly
correlated and reveals high degree of  coherence when it is
 well beyond the critical values of density and
temperature. In the framework of the approach under consideration this state
can be interpreted as a quark (constituent)-pion liquid in the nonperturbative
vacuum $V_{np}$. 

The following question arises: what one should expect at higher temperatures, e.g. at the LHC energies, i.e.
would one observe transition $V_{np}\to V_{pt}$ finally, or  other possibilities exist? Indeed, due to large 
kinetic energy of the constituent quarks in the nonperturbative vacuum
there should be a finite probability to form colorless clusters again,
i.e. confinement mechanism could take place, transition $V_{np}\to V_{ph}$ would happened instead of $V_{np}\to V_{pt}$ 
and hadrons would reappear. 
This hypothetical possibility
obtain support from unitarity saturation at very high energies. 
In the next section we  discuss this problem assuming that unitarity
is saturated at very high energies.
\section{Deconfinement and saturation of unitarity (reflective scattering)}
 The elastic scattering $S$-matrix (i.e. the $2\to 2$ scattering matrix element)
in the impact parameter representation can be
written (in the rational unitarization scheme) in the form of linear 
fractional transform (cf. \cite{inta} and references therein):
\begin{equation}
S(s,b)=\frac{1+iU(s,b)}{1-iU(s,b)}, \label{um}
\end{equation}
where $U(s,b)$ is the generalized reaction matrix. It is
considered to be an input dynamical quantity. The explicit form of the function $U(s,b)$ and numerical
predictions for the observable quantities depend on the particular  model used for hadron scattering description.
For the qualitative purposes it
is sufficient that this function increases with energy in a power-like way
 and decreases with impact parameter like a linear exponent or Gaussian \footnote{In fact, analitical properties
of the scattering amplitude imply linear exponential dependence at large values of $b$.}.
 Also for simplicity consider for the time being
 the case of pure imaginary $U$-matrix and make the replacement $U\to iU$, i.e.
\begin{equation}
S(s,b)=\frac{1-U(s,b)}{1+U(s,b)}. \label{umi}
\end{equation}
It can  easily be seen that the new scattering mode, reflective scattering (when $S(s,b)<0$) starts to appear at
the energy $s_R$, which is determined as a solution of the equation
\[
U(s_R,b=0)=1.
\]
Indeed, the unitarity relation
written for the elastic scattering amplitude $f(s,b)$  in the high
energy limit has the following form
\begin{equation}
\mbox{Im} f(s,b)=h_{el}(s,b)+h_{inel}(s,b). \label{ub}
\end{equation}
Inelastic overlap function $h_{inel}(s,b)$
is connected with $U(s,b)$ by the relation
\begin{equation}\label{hiu}
h_{inel}(s,b)=\frac{ U(s,b)}{[1+U(s,b)]^{2}},
\end{equation}
and the only condition to obey unitarity
 is $\mbox{Im} U(s,b)\geq 0$. 
Elastic overlap function is related to the function
 $U(s,b)$ as follows
\begin{equation}\label{heu}
h_{el}(s,b)=\frac{[U(s,b)]^{2}}{[1+U(s,b)]^{2}}.
\end{equation}

At sufficiently high energies inelastic overlap function
 $h_{inel}(s,b)$ would have a peripheral $b$-dependence and will
tend to zero for $b=0$ at $s\to\infty$ {cf. e.g \cite{inta}).
Therefore, corresponding behavior of elastic scattering
$S$-matrix (note that $S(s,b)=1+2if(s,b)$) can then be interpreted
as an appearance of a reflecting ability of scatterer due to increase of
 its density beyond some critical value.  In another words, the scatterer has now not only
 absorption ability (due to  presence of inelastic channels), but it starts to be reflective at very
 high energies. In central collisions, $b=0$, elastic scattering
  approaches to the completely
 reflecting limit $S=-1$ at $s\to\infty$. As a side remark, it should be noted that absorption is not a
result of imaginary nature of scattering amplitude, at small impact parameters reflective scattering
mode can exist for the pure imaginary scattering amplitude.

At the  energy values $s>s_R$ the equation $U(s,b)=1$ has a solution in the
physical region of impact parameter values, i.e. $S(s,b)=0$ at $b=R(s)$. 
The probability of reflective scattering at $b<R(s)$ and $s> s_R$ is determined by the magnitude
 of $|S(s,b)|^2$; this probability is equal to zero at $s\leq s_R$ and $b\geq R(s)$.
The dependence of $R(s)$
 is determined then by the logarithmic functional dependence $R(s) \sim \frac{1}{M}\ln s$ , this dependence
 is consistent with analytical properties of the resulting elastics scattering amplitude in
  the complex $t$-plane and mass $M$ can be related to the pion mass.
Thus, at the energies $s> s_R$  reflective scattering will mimic presence of repulsive core in
 hadron and meson interactions and elastic scattering will be dominating process. This kind of elastic scattering preserves
the hadron identities and acts against deconfinement. It would lead to the new phase of hadron liquid at very 
high temperatures. 

The following transitions can be foreseen as the temperatures increses at the constant value of chemical
potential $\mu$:
\[
 V_{ph}(\mbox{Hadron gas})\to V_{np}(\mbox{Quark-pion liquid})\to V_{ph}(\mbox{Hadron liquid}).
\]
Corresponding phase diagram depicted in Fig. 1.
\begin{figure}[hbt]
\begin{center}
\includegraphics[scale=0.4]{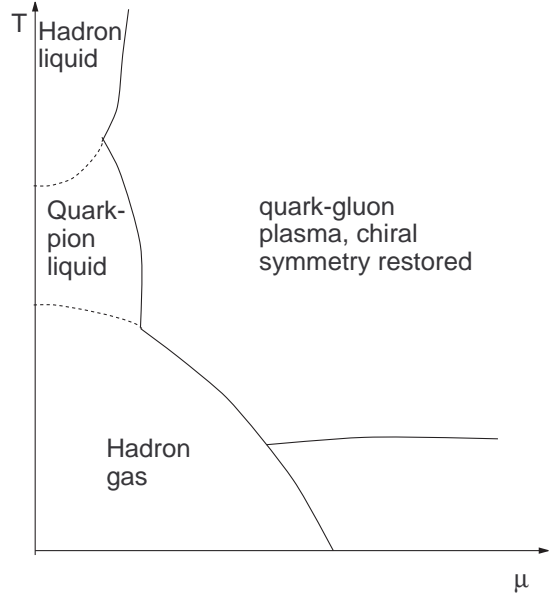}
\caption{\small\it Phases of strongly interacting matter.}
\end{center}
\end{figure}

Presence of the reflective scattering can be
 accounted for  using van der Waals method (cf. \cite{cleym}).
 This approach was originally used   for description of  fluids  starting from
 the gas approximation introducing the nonzero size of molecules into consideration.

The hadronic liquid density $n_R(T,\mu)$  can be connected \cite{limdens}
with the density in the approach without reflective scattering $n(T,\mu)$
 by the following relation
\[
  n_R(T,\mu)=\frac{n(T,\mu)}{1+\kappa(s)n(T,\mu)},
\]
where $\kappa(s)={p_R(s)V_R(s)}/{2}$, 
$p_R(s)$ is the averaged over volume $V_R(s)$ probability of reflective scattering
and the volume $V_R(s)$ is determined by the radius of the reflective scattering.
At very high energies ($s\to\infty$)
\[
n_R(T,\mu)\sim 1/\kappa(s)\sim M^3/\ln ^3 s.
\]
This  limiting dependence for the hadron liquid density  appears due to
presence of the reflective scattering which resembles in the oversimplified geometrical picture a
scattering of hard spheres in head-on hadron collisions.
 It can also be associated with  saturation of the Froissart-Martin bound for the total cross-section. 
It should be noted that the lower densities of hadron matter are needed
for percolation, but percolation in the presence of reflective scattering would not lead
to deconfinement and at very high temperatures confined phase corresponding to hadron
liquid could exist due to unitarity saturation.
\section*{Conclusion}
Thus, it was conjectured that  saturation of unitarity  would restore confinement and
percolation mechanism alone is not sufficient for deconfinement as it was supposed in \cite{satz,satz2,cleym,limdens}. In general,
we would like to note that at very high temperatures there is a certain probability that matter would return
to confined state if the unitarity saturation would occur there. 

The problem of completeness of asymptotic
colorless states deserves further disccussion \cite{rlst}. In this connection it is important to study
a possible existence of the confined phase where chiral symmetry is restored \cite{gloz} and inclusion of
such confined states into the set of asymptotic states.

Evidently, experimental studies with heavy ions at LHC would be able to reveal new  phases of matter.

\end{document}